\documentclass[journal]{IEEEtran}
\IEEEoverridecommandlockouts

\newtheorem{prop}{Proposition}

\ifCLASSINFOpdf
\else
\fi
\usepackage{xfrac}
\usepackage{mathtools,amsmath}
\usepackage{multirow}
\usepackage{color}
\usepackage{amssymb}
\usepackage{cite}

\usepackage{hyperref}
\usepackage{lipsum}
\usepackage{algpseudocode,algorithm}
\DeclareMathAlphabet{\mathpzc}{OT1}{pzc}{m}{it}
\usepackage[cal=dutchcal]{mathalfa}
\ifCLASSOPTIONcompsoc
\usepackage[caption=false,font=normalsize,labelfon
t=sf,textfont=sf]{subfig}
\else
\usepackage[caption=false,font=footnotesize]{subfi
g}
\fi

 \usepackage{etoolbox}

\usepackage{stfloats}

\begin{document}
\bstctlcite{IEEEexample:BSTcontrol}

\title{ Power-Time Channel Diversity (PTCD): A Novel Resource-Efficient Diversity Technique for 6G and Beyond}


%
%
%

\author{ Ferdi Kara,~\IEEEmembership{Senior Member,~IEEE,} Hakan Kaya, Halim Yanikomeroglu,~\IEEEmembership{Fellow,~IEEE.}
\thanks{The work of F. Kara is supported by TÜBİTAK under 2219 Postdoctoral Scholarship.}
\thanks{F. Kara is 
with the Computer Engineering, Zonguldak Bulent Ecevit University, Zonguldak, Turkey. He is also with the Department of Systems and Computer Engineering, Carleton University, Ottawa, K1S 5B6, ON, Canada  e-mail: f.kara@beun.edu.tr.}
\thanks{H. Kaya is 
with the Electrical-Electronics Engineering, Zonguldak Bulent Ecevit University, Zonguldak, Turkey,  e-mail: hakan.kaya@beun.edu.tr.}
\thanks{H. Yanikomeroglu is with the Department of Systems and Computer Engineering, Carleton University, Ottawa, K1S 5B6, ON, Canada, e-mail:halim@sce.carleton.ca.}}

\maketitle
\begin{abstract}
Diversity techniques have been applied for decades to overcome the effects of fading, which is one of the most challenging problems in wireless communications due to the randomness of the wireless channel. However, existing diversity techniques are resource-inefficient due to orthogonal resource usage, or they have high-power consumption due to multiple antennas and RF-chains which present an insurmountable constraint for small devices. To address this, this letter proposes a novel resource-efficient diversity technique called power-time channel diversity (PTCD). In PTCD, interleaved copies of the baseband symbols are transmitted simultaneously with weighted power coefficients. The PTCD provides a diversity order of the number of copies by implementing successive interference canceler at the receiver. To achieve this diversity, no additional resources are needed; hence, spectral efficient communication is guaranteed. Additionally, the power consumption at the transceivers is limited since the PTCD requires only one RF-chain. We provide an information-theoretic proof that the PTCD could have any diversity order. Based on extensive simulations, we reveal that PTCD can also outperform benchmarks without any additional cost.

\end{abstract}

\section{Introduction}
The most recent Cisco report reveals that nearly two thirds of the global population will have an Internet connection by 2023. Meanwhile, connected devices of which seventy percent are wireless, are expected to exceed the global population by over three times \cite{Cisco2020}. However, a fundamental problem of wireless communication (i.e., fading), which makes reliable communication challenging, has yet to be adequately solved. To combat the fading effect of a wireless channel, the most effective solution is to have diversity \cite{Brennan1959}. Diversity in a wireless channel simply means receiving independent (e.g., orthogonal) copies of the same signal over different channels (e.g., time, frequency, antenna). The most popular diversity technique for over thirty years has been antenna diversity \cite{Vaughan1987}, where multiple antennas are placed at the transmitter or receiver or at both ends. By having multiple antennas at the transmitter, space-time block coding (STBC) is used to enable a reliable communication and achieve a transmit diversity \cite{Tarokh1999,Tarokh1999a}. In STBC, multiple copies of the modulated base-band symbols (modified, such as conjugates and multiplied with coefficients) are transmitted with multiple antennas by using multiple time slots. The most well-known STBC is the Alamouti transmission \cite{Alamouti1998}. On the other hand, to have a receive diversity, a transmitted signal is received by multiple receiving antennas and then combined to improve signal quality \cite{Brennan1959}. The most popular method for combining in a receive diversity is maximal-ratio-combining (MRC), where each signal received by a receiving antenna is combined by multiplying the related channel conjugate \cite{Lo1999}. However, antenna diversity has some physical constraints. The multiple antennas should be placed at a distance of at least half of the wavelength. This presents an insurmountable constraint for small wireless devices. Besides, when using multiple antennas, transceivers should have multiple RF chains, which can be costly and consume a lot of power.  For newer applications, such as sensor networks and Internet-of-Things applications (according to \cite{Cisco2020}, these are expected to account for half of all connected devices), where wireless transceivers are very small and should be energy efficient, antenna diversity is not applicable. There are also other diversity techniques that can alleviate fading. In \cite{Boutros1998a}, the authors proposed a modulation diversity (a.k.a., signal space diversity) technique, where the in-phase and quadrature components of a modulated signal were interleaved to achieve a maximum diversity order of 2. Furthermore, a cooperative diversity (a.k.a., virtual multiple-input multiple-output, [MIMO]) was proposed in \cite{Sendonaris2003,Laneman2004a}, where a relay may act as an additional (virtual) antenna to guarantee a diversity order. However, in cooperative diversity, the spectral efficiency decays due to the usage of multiple hops \cite{Bletsas2006a}. Besides, cooperative diversity causes an unfairness for the relay which should consume its energy for a transmission where its own data is not included.

Based on aforementioned literature survey and discussion, existing diversity techniques that use orthogonal resources (e.g., frequency, time) are resource-inefficient. They require multiple antennas and RF chains, which necessitate large devices and a lot of power. For these reasons, these techniques may not be suitable for 6G and beyond networks. This is particularly the case for IoT applications (e.g., sensor networks), where devices may be very small and battery life limited. A resource-efficient diversity technique is therefore clearly needed to prevent fading effects in 6G and beyond applications where there are physical and energy constraints.
Motivated by this, in this letter, we propose a novel diversity technique for device-to-device communications. Our main contributions are summarized as follows.
\begin{itemize}
    \item We propose a novel resource-efficient diversity technique named power-time channel diversity (PTCD).  In PTCD, we perform coordinated interleaving (CI) and then motivated by well-known broadcast channel transmission \cite{tse_viswanath_2005, Cover2005} and its application to new multiple access schemes \cite{Ding2015_coop}, we transmit the sum of weighted interleaved copies (i.e., superposition coding) by a single antenna. At the receiver, we recover symbols by a using successive interference canceler (SIC) plus MRC of de-interleaved copies. We present a fairly simple PTCD transceiver design that uses well-known practical structures (i.e., CI, SIC, and MRC) to benefit their high performances.
    \item Thanks to the enriched analytical framework of SIC detection, we provide an information-theoretic proof for the proposed PTCD to show that it can achieve any diversity order according to the number of interleaved copies. In this regard, we derive the outage probability (OP) and diversity order of PTCD over Rayleigh fading channels.
    \item Based on extensive simulation results, we demonstrate that the proposed PTCD guarantees the target diversity. We also validate that our OP and diversity analysis for Rayleigh channel match well with simulations. Then, we present simulation results for Nakagami-$m$ fading channels and prove that the proposed PTCD always guarantees a full diversity order regardless of fading conditions.
    \item We also present comparisons with benchmark schemes (i.e., direct transmission, antenna diversity, and cooperative diversity). Based on the comparisons, we reveal that PTCD can outperform existing diversity techniques in addition to its cost-efficient and energy-efficient design.
\end{itemize}

The remainder of this letter is organized as follows. In Section II, we introduce the proposed PTCD and provide a design for the transceiver. Then, in Section III, we prove the diversity order of the proposed PTCD. The OP and diversity order analysis is also presented over Rayleigh fading channels. In Section IV, we present numerical results to evaluate the proposed PTCD and validate the analysis in comparison with benchmarks. Finally, in Section V, we conclude with a brief discussion of open research problems for PTCD.

\section{Proposed Power-Time Channel Diversity (PTCD)}
\begin{figure*}
		\centering
    \includegraphics[width=\textwidth]{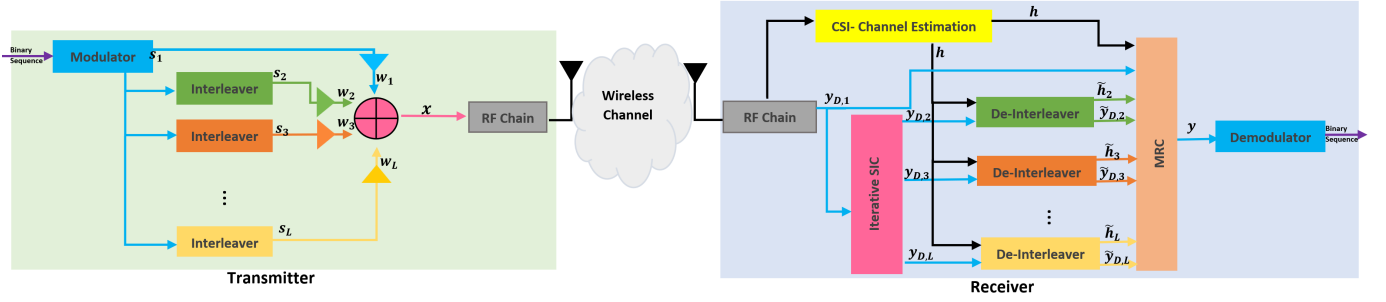}
    \caption{The proposed power-time channel diversity (PTCD) transceiver.}
    \label{PTCD_system_model}
\end{figure*}
We consider a point-to-point communication system with one source (S) and one destination (D) node. We assume that both nodes are equipped with a single antenna\footnote{Although this letter considers a single-input single-output (SISO) case, the proposed PTCD could be implemented in any antenna configuration, e.g., single-input-multiple-output (SIMO), multiple-input single-output (MISO), multiple-input multiple-output (MIMO).} and the channel fading between them is flat fading.

The transceiver design for the proposed PTCD is shown in Fig. 1 where, within a time interval $T$, the source transmits the weighed sum of its baseband symbols and interleaved copies with multiple CIs. Thus, the transmitted total symbol is given as
\begin{equation}
    x(t-kT_s)=\sum_{i=1}^{L}\sqrt{w_i}s_i(t-kTs) \quad k=1,2,\dots, \sfrac{T}{T_s},
\end{equation}
where $T_s$ is the sample period for the baseband symbols, where $\sfrac{T}{T_s}$ has an integer value. $L$ is the total number of copies (one is the output of the modulator plus $L-1$ interleaved). $s_1$ is the baseband symbol at the output of the modulator, while $s_i, \ i=2,3,\dots, L$ are the outputs of each CI. $w_i, \ i=1,2,\dots, L$ are the power weights for each copies. We assume $w_1>w_2>\dots>w_L$ (for notation simplicity) and $\sum_{i=1}^Lw_i=1$ that limits the total power consumption. Based on the transmitted symbol in (1), the received signal at the destination is given by
\begin{equation}
        y_D(t-kT_s)=\sqrt{P}x(t-kT_s)h(t-kT_s)+n(t-kT_s),
\end{equation}
where $P$ is the transmit power. $h(t-kT_s)$ is the flat fading channel coefficient, which can follow any fading model. $n(t-kT_s)$ is the additive white Gaussian noise (AWGN), which follows $CN(0, N_0)$ whose real and imaginary parts are Gaussian distributions and follow $N(0, \sfrac{N_0}{2})$.

Based on the received $y(t-kT_s), \ k=1,2,\dots, \sfrac{T}{T_s}$, we first apply pre-processing where an iterative successive interference canceler (SIC) is implemented. In the pre-processing phase, according to the order of weights, we obtain the received signal for each non-interleaved and interleaved copies. To understand this, we should note that within a time interval $T$, the received signal for the non-interleaved copy (i.e., the first branch\footnote{Note that the term ``branch" refers to each weighted copy of the superimposed transmitted signal, where the first branch is the output of the modulator (the first input of MRC) and the next $i=2,3,\dots,L$ branch is the output of the $i-1$th CI ($i=2,3,\dots,L$ inputs of MRC). See the colored lines and boxes at the transceiver in Fig. 1.} with the highest power weight) is given as
\begin{equation}
\begin{split}
        &y_{D,1}(t-kT_s)=\underbrace{\sqrt{Pw_1}s_1(t-kT_s)h(t-kT_s)}_{\text{desired signal}}+\\&\underbrace{\sum_{i=2}^L\sqrt{Pw_i}s_i(t-kT_s)h(t-kT_s}_{\text{interference due to interleaved copies}})+\underbrace{n(t-nT_s)}_{\text{noise term}},
\end{split}
\end{equation}
where $y_{D,1}(t-kT_s)\triangleq y_D(t-kT_s)$ is defined for notational simplicity. Then, after a reliable and successful SIC process, the received signal for each interleaved copy (i.e., the $i$th branch) is given by
\begin{equation}
\begin{split}
        &y_{D,i}(t-kT_s)=\underbrace{\sqrt{Pw_i}s_i(t-kT_s)h(t-kT_s)}_{\text{desired signal for the $i$th interleaved copy}}+\\&\underbrace{\sum_{j=i+1}^L\sqrt{Pw_j}s_j(t-kT_s)h(t-kT_s}_{\text{interference due to $j>i$th interleaved copies}})+\underbrace{n(t-kT_s)}_{\text{noise term}}.
\end{split}
\end{equation}
We should note that for the last interleaved branch (with the lowest power weight), this becomes
\begin{equation}
\begin{split}
   y_{D,L}(t-kTs)=\sqrt{Pw_L}s_L(t-kT_s)h(t-kT_s)+n(t-kT_s).
   \end{split}
\end{equation}
After the pre-processing steps between (3) and (5), a maximum ratio combining (MRC) is implemented, and its output is given by
\begin{equation}
\begin{split}
        y(t-kT_s)=&y_{D,1}(t-kT_s)h^*(t-kT_s)\\&+\sum_{i=2}^L\tilde{y}_{D,i}(t-kT_s)\tilde{h}^*(t-kT_s),
\end{split}
\end{equation}
where $(\cdot)^*$ is the complex conjugate operation. As shown in Fig. 1, $\tilde{y}_{D,i}$ denotes the de-interleaved received signal for the $i$th branch. Likewise, $\tilde{h}_{i}$ is the de-interleaved channel response according to the $i$th interleaver.

\section{Performance Analysis}
\subsection{Signal-to-Interleaved-Interference  Plus Noise Ratio (SIINR) Definitions}
As presented in (3)--(4), in each step of the iterative SIC, the received signal from each branch is exposed to self-interference due to the sum of interleaved copies. Thus, the signal-to-interleaved interference plus noise ratio (SIINR) for $i=1,2,\dots, L-1$ is given by
\begin{equation}
    SIINR_i=\frac{w_i|h(t-kT_s)|^2}{\sum_{j=i+1}^Lw_j|h(t-kT_s)|^2+\frac{1}{\rho}},
\end{equation}
where $\rho=\sfrac{P}{N_0}$ is defined. And, according to (5), the SIINR for the $L$th branch becomes
\begin{equation}
    SIINR_L=\rho w_L|h(t-kT_s)|^2.
\end{equation}

After the SIC, a coordinated de-interleaving is applied for $2\leq i\leq L$ branches. After the de-interleaving, the SIINR definitions in (7)--(8) turn out to be
\begin{equation}
    \tilde{SIINR}_i=\frac{w_i|\tilde{h}_i(t-kT_s)|^2}{\sum_{j=i+1}^Lw_j|\tilde{h}_i(t-kT_s)|^2+\frac{1}{\rho}}, \ i=2,\dots, L-1,
\end{equation}
and
\begin{equation}
    \tilde{SIINR}_L=\rho w_L|\tilde{h}_i(t-kT_s)|^2.
\end{equation}
Finally, since an MRC is applied as given in (6), the total SIINR at the output of the detector is given as
\begin{equation}
    \begin{split}
        SIINR&=SIINR_1+\sum_{i=2}^L\tilde{SIINR}_i\\
        &=\frac{w_1|h(t-kT_s)|^2}{\sum_{j=2}^Lw_j|h(t-kT_s)|^2+\frac{1}{\rho}}\\&+\sum_{i=2}^{L-1}\frac{w_i|\tilde{h}_i(t-kT_s)|^2}{\sum_{j=i+1}^Lw_j|\tilde{h}_i(t-kT_s)|^2+\frac{1}{\rho}}\\&+\rho w_L|\tilde{h}_i(t-kT_s)|^2.
    \end{split}
\end{equation}
For simplicity of notation, after this point, we use $SIINR_i\triangleq \tilde{SIINR}_i$ for $i\leq2\leq L$.
\subsection{Outage Probability}
The outage occurs when the quality-of-service (QoS) requirement is not satisfied. Thus, the outage probability is defined as
\begin{equation}
    P(out)=P(SIINR<\gamma_{th}),
\end{equation}
where $\gamma_{th}=2^{\acute{R}}-1$ and $\acute{R}$ denotes the QoS requirement (i.e., bit per channel usage [BPCU]). By substituting (11) into (12), we obtain
\begin{equation}
    P(out)=P(\sum_{i=1}^LSIINR_i<\gamma_{th}).
\end{equation}
\begin{prop}
The OP in (13) is upper-bounded by
\begin{equation}
    P(out)=P(\sum_{i=1}^LSIINR_i<\gamma_{th})\leq\prod_{i=1}^LP(SINR_i<\gamma_{th}).
\end{equation}
\end{prop}
\begin{IEEEproof}
Recall that at the transmitter side within a $T$ time interval, the $s(t-kT_s)$ baseband signal is conveyed to the destination with $w_1$ weight by a direct path (no interleaving). But, thanks to the $L-1$ branch interleaving, it is also transmitted with a $w_i, \ i=2,3,\dots, L$ weight by the $i$th interleaved branch by $s(t-p_iT_s)$. In addition, thanks to the different interleaving in each branch, we guarantee that $k\neq p_i,\ \forall  i$ and $p_i \neq p_j,\ \forall  \ i \neq j$. This guarantees that any symbol  $s(t)$ is received at the destination with different $w_i$ weights through different channel coefficients. Then, if we look at the $SIINR_i$ definitions in (7), (9), and (10), we can see that they are functions of the channel coefficient $h(t-kT_s)$ and de-interleaved $\tilde{h}_i(t-kT_s), \ i=2,3,\dots, L$, which are independent coefficients in any $T$ time interval.

Recall that for independent $\theta$ and $\phi$ events, their probabilities are also independent. Thus, $P(\theta+\phi<\zeta)\leq P(\theta<\zeta).P(\phi<\zeta)$. Applying this for $L$ branch-independent coefficients, the upper bound is given as in (14). So the proof is completed.
\end{IEEEproof}
\subsubsection{Closed-form OP upper-bound for Rayleigh fading channels}
We assume that the channel coefficient $h$ (or $\tilde{h}_i, \ \forall i$) follows $CN(0,1)$, so that $|h|^2$ follows an exponential distribution. Thus, $P(SIINR_i<\gamma_{th})$ is given by
\begin{equation}
\begin{split}
    &P(SIINR_i<\gamma_{th})=\\
    &\begin{cases} 1, &\text{if}\ \gamma_{th}>\frac{w_i}{\sum_{j=i+1}^Lw_j},\\
    1-e^{-\frac{\gamma_{th}}{\rho\left(w_i-\sum_{j=i+1}^Lw_j\gamma_{th}\right)}}, &\text{otherwise}.
    \end{cases}
\end{split}
\end{equation}
Hence, the OP of the proposed PTCD is upper bounded by
\begin{equation}
\begin{split}
    &P(out)\leq\\
    &\begin{cases} 1, &\text{if}\ \gamma_{th}>\frac{w_i}{\sum_{j=i+1}^Lw_j},\\
    \prod_{i=1}^L\left(1-e^{-\frac{\gamma_{th}}{\rho\left(w_i-\sum_{j=i+1}^Lw_j\gamma_{th}\right)}}\right), &\text{otherwise}.
    \end{cases}
    \end{split}
\end{equation}
\subsection{Diversity Analysis}
The diversity order of the PTCD is given by
\begin{equation}
    \delta=-\lim_{\rho\rightarrow\infty}\frac{10\log P(out)}{10\log\rho}.
\end{equation}
\begin{prop}
As long as the  $\gamma_{th}>\frac{w_i}{\sum_{j=i+1}^Lw_j}, \ \forall i, \ i=1,2,\dots,L$ condition is satisfied, the proposed PTCD guarantees a diversity order of $L$.
\end{prop}
\begin{IEEEproof}
Firstly, we define the $P(SIINR_m<\gamma_{th})=\max\{P(SIINR_i\gamma_{th})\}, \ i=1,2,\dots, L$. In this case, the upper-bound in (14) can be given as (relaxed)
\begin{equation}
    P(out)\leq\left(P(SIINR_m<\gamma_{th})\right)^L.
\end{equation}
By substituting (18) into (17), we obtain
\begin{equation}
\begin{split}
        \delta=&-\lim_{\rho\rightarrow\infty}\frac{10\log\left(P(SIINR_m<\gamma_{th})\right)^L}{10\log\rho}\\&=-\lim_{\rho\rightarrow\infty}L \frac{10\log P(SIINR_m<\gamma_{th})}{10\log\rho}=L\cdot
\end{split}
\end{equation}
The proof is completed.
\end{IEEEproof}

Here, we note that the above diversity order of $L$ is guaranteed for any fading channel. According to the characteristics of the fading channel, the diversity order $L$ is multiplied by the channel parameter. For instance, the diversity order of PTCD is given by $L$ for a Rayleigh fading channel, whereas it is equal to $mL$ for a Nakagami-$m$ fading channel.


\section{Numerical Results}
In this section, we present the results of computer simulations to evaluate the proposed PTCD. With the simulations, we also validate the OP and diversity analysis provided in the previous section. In all simulations, we use the following power weights:\footnote{We should note that by optimizing the weights, the performance of the PTCD can be further improved horizontally (like the coding gain in STBC). However, weight optimization is beyond the scope of this letter and will be considered future work.} $\mathbf{w}=[0.8,\ 0.2]$, $\mathbf{w}=[0.9,\ 0.09,\ 0.01]$,  $\mathbf{w}=[0.8,\ 0.15, \ 0.04, \ 0.01]$ for $L=2,\ 3,\ 4$ where $\mathbf{w}\triangleq[w_1,\ w_2,\ \dots,\ w_L]$. In addition, unless otherwise stated, the QoS requirement is equal to $\acute{R}=1$ BPCU.
We also provide comparisons with three benchmarks to evaluate the performance of the proposed PTCD. The simulation set-ups for the benchmarks are given below.

\textit{i) Direct transmission (no diversity)}: In the first benchmark, we assume that no diversity path is available. Hence, a point-to-point communication is presented with a single transmit and receive antenna at the transceivers. For fairness in comparisons, the transmit power of the direct transmission is equal to the total transmit power of the PTCD and the QoS requirement of direct transmission remains the same ($\acute{R}_{direct}=\acute{R}$).

\textit{ii) Space-time-block-coding (STBC) diversity}: In STBC, we consider multiple antennas located at the transmitter. Copies of consecutive symbols of the destination (and variations, such as complex conjugate with different power coefficients) are conveyed to the destination through multiple transmit antennas ($T_x$) by using STBC code designs given in \cite{Tarokh1999}. For a special case, when $T_x=2$, we implement the Alamouti scheme in \cite{Alamouti1998}. The OP simulations of STBC designs are  performed by following the definitions in \cite{Chen2010}. We should note that in STBC, multiple time slots are required to convey symbols. Thus, for fairness, the QoS requirements in STBC are given as $\acute{R}_{STBC}=\sfrac{\acute{R}}{R_k}$, where $R_k$ is the code rate in STBC, which is given as $R_k=1$ for $T_x=2$ \cite{Alamouti1998} and $R_k=\sfrac{3}{4}$ for $T_x=3,\ 4$ \cite{Tarokh1999}. The same amount of power is allocated to transmit symbols in both STBC and PTCD, where we ignore the power consumption of multiple RF-chains in STBC, although it is vital for power and cost efficiencies of the devices.

\textit{ii) Cooperative Diversity}: In cooperative diversity, we assume that one direct path is available between source and destination. Besides, $L-1$ intermediate relays are located between the source and destination with a single antenna \cite{coop_book}. We assume a symmetric network \cite{Onat2008a}, where all links (i.e., between source-to-relays, relays-to-destination, and source-to-destination) undergo the same channel statistics. The relays operate in half-duplex and apply a decode-and-forward protocol, where, in the first time slot, all relays and the destination receive the source signal. Each relay then decodes the signal and forwards it to the destination successively after re-encoding it in the next $L-1$ time slots. The destination then implements an MRC to combine the received signals within $L$ time slots \cite{coop_book}.  In the comparisons, for fairness,  the total transmit power in cooperative diversity (at the source plus relays) is equal to the transmit power of PTCD. Again, we ignore the power consumption of RF-chains in each relay. Furthermore, the QoS requirement for cooperative diversity is given by $\acute{R}_{coop}=\acute{R}L$ since $L$ time slots are required to transmit one symbol in cooperative diversity.

In Fig. 2, we present the outage performance of the proposed PTCD for various numbers of interleaving branches over a Rayleigh fading channel. First, we can see that the derived upper bound OP expression matches well with the simulations, which proves the correctness of the analysis and effectiveness of the proposed PTCD. Additionally, the proposed PTCD provides an $L$ diversity order (i.e., slope of the OP curves on a logarithmic scale as given in (19)). To further investigate, in Fig. 3, we present numerical results for the diversity order over Rayleigh fading channels with $\acute{R}=0.5$ BPCU. As we can see, the diversity order of the system converges to $L$ as SNR increases, which shows that the PTCD can achieve the full diversity order.

\begin{figure}
		\centering
    \includegraphics[width=.85\columnwidth]{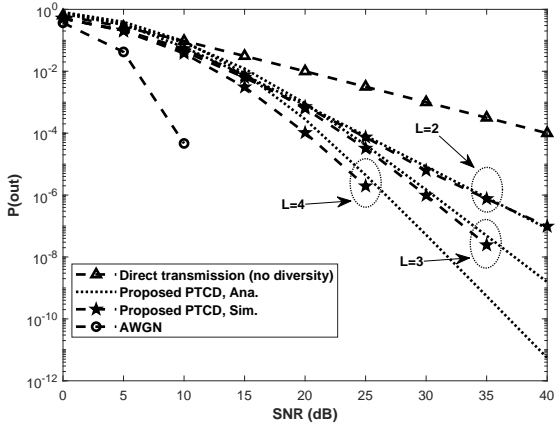}
    \caption{Outage performance of PTCD over Rayleigh fading channel.}
    \label{OP_rayleigh}
\end{figure}

\begin{figure}
		\centering
    \includegraphics[width=.85\columnwidth]{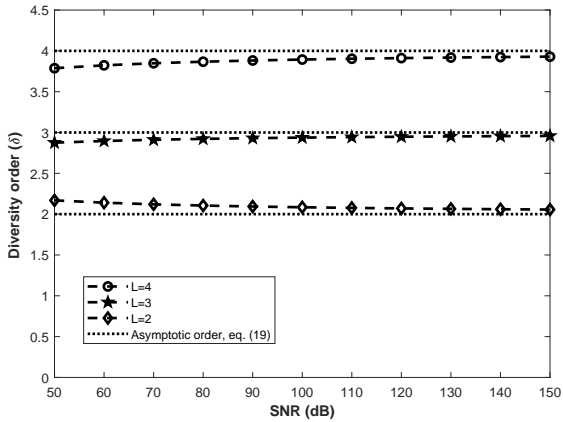}
    \caption{Diversity of PTCD over Rayleigh fading channel.}
    \label{diversity_rayleigh}
\end{figure}


Fig. 4 presents comparisons with three benchmarks in various configurations. Regardless of the $L$ (i.e., number of copies in PTCD and number of total branch [one direct path plus $L-1$ relaying path] in cooperative diversity), the PTCD outperforms the cooperative diversity in all scenarios. Yet we can see that the STBC schemes appear to perform better than the proposed STBC. It is worth noting that, in STBC, the required RF-chain increases with the number of transmit antennas. This causes high power consumption at the transceivers in addition to the physical constraint of mounting multiple antennas. This can be particularly challenging when connected devices are very small (e.g., IoT devices and sensor networks). In addition, we should note that PTCD can easily be used with other diversity techniques. For example, it can also be implemented when STBC is used to increase the reliability of the communication.

\begin{figure}
		\centering
    \includegraphics[width=.85\columnwidth]{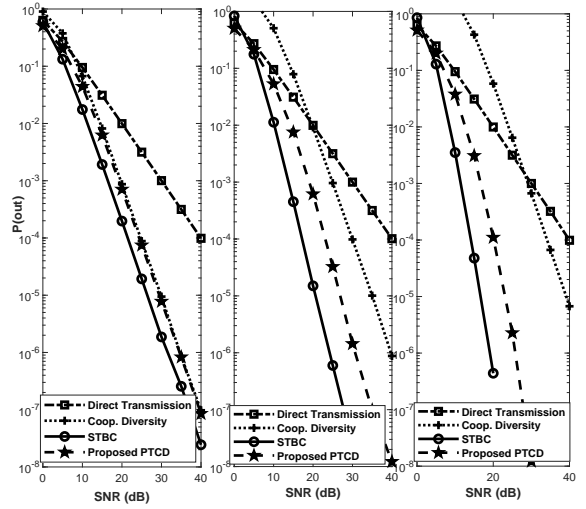}
    \caption{Outage performance comparisons for cooperative diversity, STBC and proposed PTCD. a) $L=2$ b) $L=3$ c) $L=4$.}
    \label{comparisons}
\end{figure}

Finally, to demonstrate the robustness of PTCD in different fading environments, in Fig. 5, we present the OP performance of PTCD over Nakagami-$m$ fading channels, where the spread parameter is normalized. Fig. 5 also shows the results for various shape parameters ($m$), where we consider both worse (i.e., $m=0.5$) and better (i.e., $m=1.5,\ 2$) channel conditions than Rayleigh fading.  As we can see, PTCD provides a full diversity order of $mL$ regardless of the number of interleaved copies ($L$) or shape parameters of the channel ($m$). This proves that PTCD provides an effective diversity in various channel conditions.

\begin{figure}
		\centering
    \includegraphics[width=.85\columnwidth]{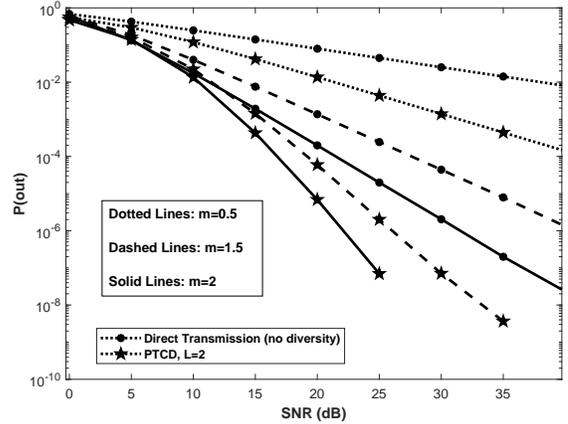}
    \caption{Outage performance of PTCD over Nakagami-$m$ fading channel.}
    \label{OP_nakagami}
\end{figure}

\section{Conclusion}
In this letter, we proposed a novel diversity technique named power-time channel diversity (PTCD). As a part of this technique, we presented a transceiver design based on coordinated interleaving and superposition coding at the transmitter, and an iterative successive interference canceler at the receiver. Then, we provided an OP analysis and proved that PTCD can achieve a full diversity order ($L$). Based on the extensive simulations, we demonstrated that PTCD can outperform benchmark diversity techniques and work well in various fading conditions. While this letter dealt with the design of PTCD and its fundamental limits, the performance of PTCD can further be improved with a weight optimization. Future studies will address this aspect of PTCD. Furthermore, PTCD can also be implemented in existing schemes. The study of such implementations is also a promising direction of future research.

%
\ifCLASSOPTIONcaptionsoff
  \newpage
\fi
%
\bibliographystyle{IEEEtran}

\bibliography{kara_WCL2022_0653}
%
\end{document}